\documentclass{aa}

\newcommand{\grp}    {${\rlap.}^{\circ}$}
\newcommand{\pri}    {${\rlap.}^{\prime \prime}$}

\newcommand{\ltsima} {$\; \buildrel < \over \sim \;$}
\newcommand{\simlt}  {\lower.5ex\hbox{\ltsima}}            
\newcommand{\gtsima} {$\; \buildrel > \over \sim \;$}
\newcommand{\simgt}  {\lower.5ex\hbox{\gtsima}}            

\usepackage{graphicx}
\usepackage{txfonts}
\begin{document}

\title{The radio jets of SS~433 at millimetre wavelengths}


   \author{Josep Mart\'{\i}\inst{1}
          \and
          Irene Bujalance-Fern\'andez\inst{1}
                    \and
          Pedro L. Luque-Escamilla\inst{2}
          \and
          Estrella S\'anchez-Ayaso\inst{1}
          \and
          Josep M. Paredes\inst{3}
          \and
          Marc Rib\'o\inst{3}
                    }

  \institute{Departamento de F\'isica (EPSJ), Universidad de Ja\'en, Campus Las Lagunillas s/n Ed. A3 Ja\'en, Spain, 23071\\
\email{jmarti@ujaen.es}
\and
 Departamento de Ingenier\'{\i}a Mec\'anica y Minera (EPSJ), Universidad de Ja\'en, Campus Las Lagunillas s/n Ed. A3 Ja\'en, Spain, 23071\\
\email{peter@ujaen.es}
\and
  Departament de F\'{\i}sica Qu\`antica i Astrof\'{\i}sica,  Institut de Ci\`encies del Cosmos, Universitat de Barcelona, IEEC-UB, Mart\'{\i} i Franqu\`es 1, E-08028 Barcelona, Spain\\
}

   \date{Received XXXXXX XX, 2018; accepted XXXXXX XX, 2018}

 
  \abstract
   {SS~433 is historically a well-known microquasar in the Galaxy that has been deeply studied during the four decades elapsed since its discovery.
   However, observations at very high radio frequencies 
   with good angular resolution are still very scarce in the literature. 
   The present paper tries to 
    partially fill this gap using archival data of the source obtained with
   the Atacama Large Millimeter Array (ALMA).}
   {We aim to study the SS~433 jet properties at radio frequencies corresponding to millimetre wavelengths
    where the synchrotron emitting particles are expected to lose their energy much faster
   than at  lower frequencies of  centimetre wavelengths. }
   {We applied the methods of connected radio interferometry adapted to the ALMA spectral domain 
   for data calibration and reduction. Afterwards, the resulting map was corrected for relativistic and geometric effects. While a detailed theoretical  modelling  was beyond our scope, 
   we assessed the broad evolutive properties of the jet brightness profile
   in the context of a semi-analytic model.
 }
   {A resolved view of the SS~433 radio core and jets is presented. In addition to spectral index and magnetic field measurements, we are able to estimate the age of the oldest visible ejecta
   still radiating significantly at millimetre wavelengths. By combining our findings with those of previous authors at lower frequencies, we confirm that the energy loss
   of the radiating electrons is dominated by adiabatic expansion instead of synchrotron radiative losses. In addition,  we  find suggestive 
   evidence for the previously proposed period of slowed expansion  within the first
   months of the ejecta flow, needed to simultaneously match the radiative lifetime observed in the centimetre domain.
   Our results argue for 
   the need for future coordinated millimetre and centimetre interferometric observations with good time sampling  throughout the SS~433 precessional cycle  
   to better understand energetic processes in stellar relativistic jets.}
   {}


   \maketitle
%

\section{Introduction} 
 
Object \#433 in the \citet{1977ApJS...33..459S} catalogue (\object{SS~433}, also known as \object{V1343 Aql}) was
discovered in the late 1970s as a very peculiar X-ray and radio star
with moving emission lines in its optical spectrum \citep{1978Natur.276...44C, 1979ApJ...230L..41M}. The  circumstances of its discovery and the building of the currently widely accepted precessing jet model were first reviewed in \citet{1984ARA&A..22..507M}. Other in-depth reviews followed
decades after (e.g.  \citealt{2004ASPRv..12....1F}).
 Believed to be a unique object during the first years after its identification, today we understand \object{SS~433} to be one of the most relevant
members in the microquasar family of relativistic jet sources of stellar origin \citep{1999ARA&A..37..409M, 2016LNP...905...65F}.
This group also has remarkable implications in modern  high energy
astrophysics \citep{2005Ap&SS.300..267P, 2015ApJ...807L...8B, 2018A&A...612A..14M}.
The number of SS~433-related references in the SIMBAD Astronomical Database amounts to nearly two thousand publications  to date,
and  SS~433-dedicated books have even been published in the past \citep{clark1986quest}.

While SS~433 has been deeply studied in almost the entire electromagnetic spectrum, observations in the
millimetre domain are still not very abundant in the literature. Beyond simple flux density measurements (as in  \citealt{1989ApJ...338..945B, 2000A&A...357..507P})
to molecular emission line studies of the environment (as in \citealt{2007MNRAS.381..881L}), SS~433 has seldom been explored with
high angular resolution and sensitivity at millimetre wavelengths. This is in contrast with the abundant centimetre observations,
with different  interferometers, where the collimated bipolar jets and their proper motions at about a quarter of the speed of light are
clearly seen \citep{1993A&A...270..177V, 1999A&A...348..910P, 2001ApJ...562L..79B, 2004ApJ...616L.159B}.

In this paper we take advantage of the superb ALMA capabilities 
 by  analysing archival data of SS~433. Observing at radio frequencies well above 100 GHz allows us to sample
the non-thermal emission of relativistic electrons higher in their energy power-law spectrum. In particular, here  to investigate the
properties of the intrinsic brightness of the SS~433 jets in this frequency regime for the first time and compare it with previous studies performed in the 
most traditional bands of radio astronomy
\citep{2010ApJ...719.1918R, 2011ApJ...736..118B}. ALMA allows us to do so with excellent subarcsecond angular
resolution rivaling that achieved by the preceding generation of connected interferometers.

\section{Archival observations with ALMA}

We retrieved the only existing set of SS~433 observations in the ALMA archive that were conducted  
in the context of multi-frequency monitoring of this microquasar\footnote{Project Id. ADS/JAO.ALMA\#2013.1.01369.S, PI K. Blundell.}.
Three consecutive scheduling blocks spanning $1^{\rm h}$ each were executed starting on 2015 September 28 at $21^h 50^m 06^s$ UTC, and ending on
2015 September 29 at $01^h 40^m 22^s$ UTC.
The instrumental set-up included four spectral windows within  ALMA Band 6  centred at frequencies $\nu= 224, 226, 240,$ and 242 GHz.
Each spectral window  covered  a $\Delta \nu =2$ GHz bandwidth split into 64 channels.
The corresponding central wavelengths were in the $\lambda=$1.24--1.34 mm range.
Correlator products of XX-, YY-, and XY-type were sampled with an integration time of 2 s.
A total of 31 antennas in the C34-7 configuration were available. This provided baselines between 42 and 1574 m, equivalent
to 25 and 1200 k$\lambda$, respectively. This array set-up enabled observations with subarcsecond angular resolution and
sensitivity to angular scales as large as $\sim4^{\prime\prime}$.

Weather conditions were good and the system temperatures of most of the  antennas used were not in excess of 100 K. 
The source J1751+0939 was observed as a bandpass and polarization calibrator, while J1832+0731, 
and occasionally J1830+0619, both within $10^{\circ}$ of the target, were
observed as phase calibrators. The flux density scale was tied to observations of the asteroid Pallas. Its accuracy in Band 6 is believed to be 
in the 5\% -- 10\% range.
Calibration  was conducted using the CASA 5.1.1-5 software package. All steps in the python scripts for calibration provided by the ALMA archive
were run, and their result carefully scrutinized for improvement whenever possible by excluding bad visibilities not automatically flagged.
The three calibrated measurement sets were finally concatenated into a single set using the CASA task {\tt concat} for imaging.

\begin{figure}
   \centering
   \includegraphics[width=\hsize]{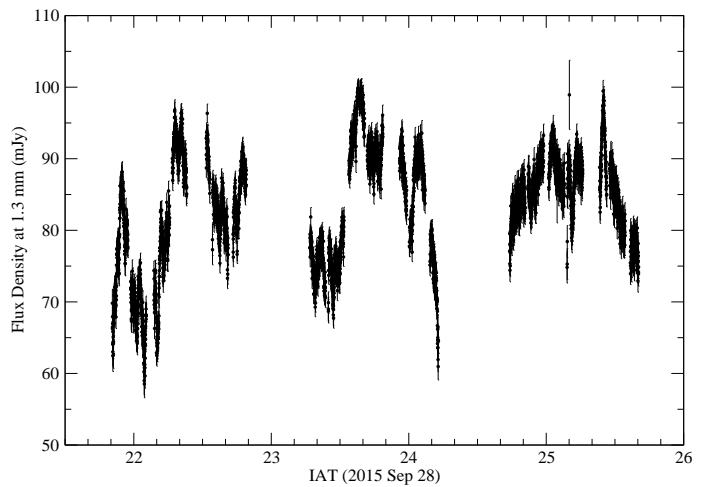}
      \caption{Radio light curve of the SS~433 central core  on 2015 September 28 and 29.
      This plot corresponds  to the 224 GHz frequency (spectral window 0 of the ALMA data).
      The time resolution is 2 s.
              }
         \label{lcurve}
   \end{figure}

\begin{figure}
   \centering
   \includegraphics[width=10.0cm]{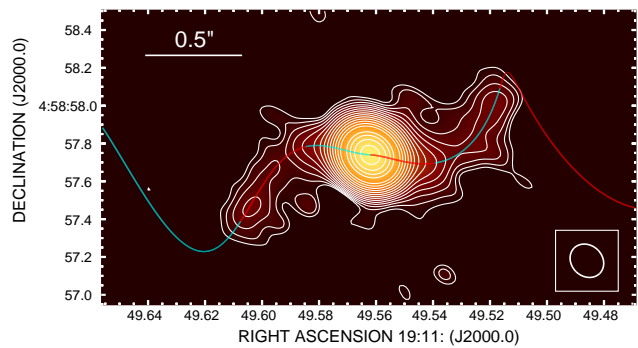}
      \caption{ALMA map of SS~433 obtained on 2015 September 28 at the 1.3 mm wavelength and computed using natural weight.
      The restoring beam used corresponds to an elliptical Gaussian with full width at half maximum
      of 190 $\times$ 161 mas$^2$, with position angle of 44\grp 8  (see bottom right corner).
      The horizontal bar gives the angular scale, with north being up and east  left.
      The jet kinematical model is overplotted in blue and red  for  
      approaching and receding material, respectively.
      Contour levels start at four times the rms noise of 32 $\mu$Jy beam$^{-1}$ and increase progressively by a factor of $\sqrt{2}$.
              }
         \label{map}
   \end{figure}

The calibrated data were also exported into the AIPS software package of NRAO. The reason for that  was to take
advantage of  the AIPS task {\tt DFTPL} intended to easily generate
light curves of stellar sources for radio photometry.  The behaviour of flux density $S_{\nu}$ 
as a function of time for the SS~433 central core is displayed in Fig. \ref{lcurve}. Flaring events
with amplitude 
 \simlt 20 mJy are superposed onto 
 an average value of about 80 mJy, and evolving on timescales of $\sim 10$ min.
The variable nature of SS~433 implies that self-calibration with a constant radio source model is not strictly correct.
The mitigation of possible unwanted effects by subtracting a central component with time-dependent flux density is addressed in Sect. \ref{iprof}.
The observed amplitude of variability
was 
not dramatic
most of the time  (within 20\%). Self-calibrating in phase and amplitude  still worked in an acceptable way despite the
variability issue, and improved 
the root-mean-square (rms) noise by nearly a factor of two in the Stokes $I$ parameter. Maps deconvolved with the CLEAN algorithm were computed  using both CASA and AIPS,
yielding very similar results. The map shown in Fig. \ref{map} was obtained using natural weight and, to our knowledge, it provides the first
subarcsecond view of the microquasar jets at the 1.3 mm wavelength. Polarization maps were also obtained for $Q$ and $U$ Stokes parameters.
However, linear polarization was only detected for the central core  at a low $7 \sigma$ level of  $\sim 0.5$\%, which will not be discussed further in this work.

Measuring the SS~433 spectral index $\alpha$ ($S_{\nu} \propto \nu^{\alpha}$) was attempted by averaging the first and second spectral window pairs,
and creating maps at two slightly different frequencies (about 225 and 241 GHz, respectively). Given the proximity of the frequencies being combined, 
the resulting spectral index map  is only reliable in the vicinity
of the bright central core where a nearly flat spectrum radio core is clearly observed ($\alpha = -0.01 \pm 0.01$). 
Hints of steepening to negative, non-thermal
values are present towards the jets where synchrotron emission is expected to become optically thin. 
Unfortunately, our limited frequency coverage does not allow us a more
precise measurement  of $\alpha$ along the fainter jet flow regions.


%

%

%
 
\section{Kinematic model}

The expected path of the SS~433 jets has been overplotted  onto Fig. \ref{map} 
for the date of the observation.
The  paths when the jet is approaching and receding from the observer are always shown in blue and red, respectively.
The kinematic model used is based on the \citet{1981ApJ...246L.141H} formalism updated with the precession parameters
derived by \citet{ 2001ApJ...561.1027E}. Their values are based on 20 years of optical spectroscopy
measuring the  Doppler motion of the jet moving lines. We must note, however, that the zero-phase given by these authors actually
corresponds to a precessional phase value of 0.33 because of a phase offset propagated in different papers.
This phase origin issue apparently goes back to the  years immediately after  the discovery of  SS~433.
A warning about it can already be found in Section II of \citet{1980ApJ...241..306M}.
In agreement with \citet{2004ASPRv..12....1F}, the adopted zero precessional phase in this paper corresponds to the Julian Day
 JD2443507.47 (29.97 December 1977).
 Nutation is not taken into account here for reasons of  simplicity  given that only one ALMA observing epoch is available.
 In addition, the induced angular displacements at $\sim 1^{\prime\prime}$ from the core are still hard to see at the ALMA angular resolution.

The precession axis position angle cannot be determined spectroscopically.  
We use the historical value of 100\grp 0 from the north to the east based on the Very Large Array observations by
 \citet{1981ApJ...246L.141H}, whose angular resolution was closer to ours than
other similar estimates from Very Long Baseline Interferometry \citep{2002MNRAS.337..657S}.
Moreover, we adopt here the canonical distance of 5.5 kpc \citep{2007MNRAS.381..881L}.
We note that this value remains consistent with the still uncertain trigonometric parallax of SS~433 recently reported in the latest {\it Gaia} 
data release  \citep{2016A&A...595A...1G}, which  implies a distance of $4.6^{+1.9}_{-1.0}$ kpc (error quoted is $\pm 1\sigma$).
The kinematic model is also used to calculate the ejection angles $\theta$ with the line of sight. These values are needed in Sect. \ref{iprof}
to compute the Doppler factor corrections for each jet segment. 
  
\section{Observed brightness profile}
 
 The observed profile is determined by exploring the FITS file of the interferometric map
along the path given by the kinematic model. For each individual path position, we store the pixel
values within a small radius (typically $\leq 2$ pixel) and fit a least-squares plane to them. 
The fitted value is taken as the observed brightness level at this position. The uncertainty associated
with it has three different contributions \citep{2011ApJ...736..118B}: the rms noise level in the image (32 $\mu$Jy beam$^{-1}$),
the overall calibration flux error ($\sim5$\% for ALMA), and the jet path position error estimated assuming a $\pm10$\% variation
in the jet velocity. All these contribution are later added in quadrature. 

\begin{figure}
   \centering
   \includegraphics[width=\hsize]{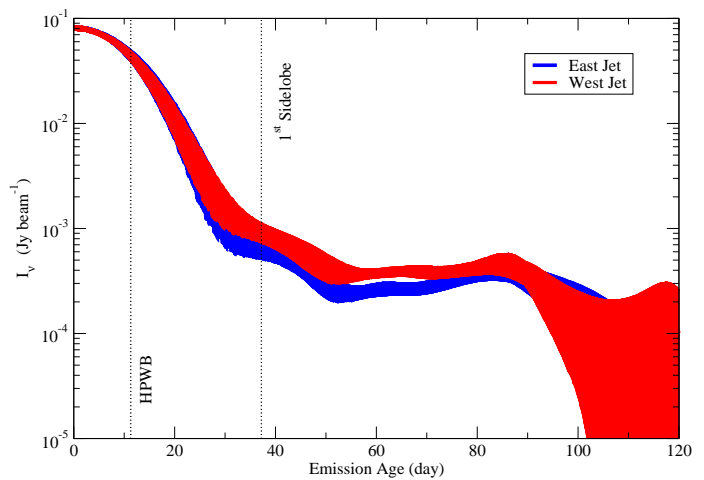}
      \caption{Observed  brightness profile of the SS~433 jets on 2015 September 28 at the 1.3 mm wavelength.
      The uncorrected jet brightness values are plotted as a function of the emission age 
      across the jet flow. 
      The  coloured bands  outline the regions of uncertainty estimated as explained in the text.
      The vertical  bars roughly correspond to the times needed by a young emerging blob to travel the half-power beam width of
       one synthesized beam and its first sidelobe. Blue and red  are always used for the east and west jet sides, respectively.
      }
               \label{profile_obs}
   \end{figure}

The outcome of this procedure applied to the Fig. \ref{map} is displayed in Fig. \ref{profile_obs}.
Any ALMA map is produced by photons emitted  from different pieces of the jet that simultaneously reach the observer.
Each of these pieces may be identified by its  ejection age $t_{\rm age}$, which is the time elapsed in the reference frame of the core
since the jet material was ejected. However, due to light-travel effects, any of the pieces is seen as it was at an emission age $\tau$ that  relates to $t_{\rm age}$ according to
\begin{equation}
\tau = \frac{t_{\rm age}}{1-\beta \cos{\theta}}, \label{ttau}
\end{equation}
where $\beta$ is the jet flow velocity in units of the speed of light $c$ and $\theta$ is the local jet angle with the line of sight predicted
by the kinematic model (see Appendix A in \citet{2010ApJ...719.1918R} for further details). The profile in Fig. \ref{profile_obs} is given as a function of $\tau$ for both the east and west jets.

To  be cautious and to avoid over-interpretation too close to our threshold  in angular resolution, we  constrain our analysis to features clearly
separated from the central core by at least one synthesized beam up to its first sidelobe.
In our case, this secondary feature in the ALMA point spread function has a 310 milliarcsecond (mas) radius with
a response still at the 10\% level.
Both the half-power beam width of the synthesized beam and its first sidelobe location are indicated as vertical bars in Figure \ref{profile_obs} 
and other similar plots in this work.
Thus, we consider that a reliable analysis of the jet brightness profile with ALMA is feasible for emission 
 ages older than about 40 days.

%

\section{Intrinsic brightness profile} \label{iprof}

Here we adapt the prescription developed in \citet{2010ApJ...719.1918R}  and \citet{2011ApJ...736..118B}
 to transform the observed brightness profile of the SS~433 jets
into the intrinsic profile corrected for projection and Doppler boosting effects. Instead of a circular synthesized beam,
an elliptical Gaussian shape is considered for slightly better accuracy with $b_{\rm maj}$,  $b_{\rm min}$, and $PA$ 
being its major axis, minor axis, and position angle, respectively.
The  correction factor needs to be computed for each
sky position $(x_i,y_i)$ along the jet path consisting of  $i=1,...,N$ segments or steps. 
Its  value results from the  position average of the
 Doppler factor $D_i^{n-\alpha} = \left[1 /\Gamma(1-\beta \cos{\theta_i})\right]^{(n-\alpha)}$ weighted by the synthesized beam of the interferometer. 
 As usual, the Lorentz factor is $\Gamma=1/\sqrt{1-\beta^2}$, and we adopt $n=2$ corresponding to a continuous jet flow.
 For the spectral index, we use the average value $\alpha=-0.74$ for the extended jets taken from \citet{2011ApJ...736..118B} as we cannot measure
 it directly beyond the SS~433 core.

   \begin{figure}
   \centering
   \includegraphics[width=\hsize]{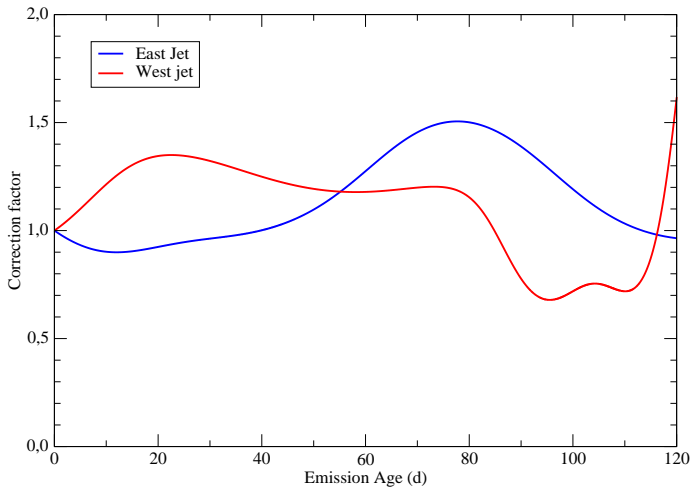}
      \caption{Multiplicative correction factors  $1/C_i$ to recover the intrinsic jet brightness profiles from the observed ones on 2015 September 28.
      They are plotted as a function of the emission age across the east and west jet flows, shown in blue and
      red, respectively.  
      The point spread function assumed is the same
      elliptical Gaussian (190 $\times$ 161 mas$^2$, with position angle of 44\grp 8) 
      used as a restoring beam in  Fig. \ref{map}. 
      }
              \label{Cn}
   \end{figure}
%
 
 Let the angular separation between the $i$-th sky position and any other jet segment be
 $\Delta x_j = x_j - x_i$  and $\Delta y_j = y_j - y_i$, where $j$ also runs from 1 to $N$.
By performing a rotation towards the main axes of the Gaussian, the $i$-th angular offsets become
$$\Delta x_j^{\prime} = \cos{(PA)} \Delta x_j - \sin{(PA)} \Delta y_j,   $$
$$\Delta y_j^{\prime} = \sin{(PA)} \Delta x_j + \cos{(PA)} \Delta y_j.  $$
The multiplicative correction factor is then
 \begin{equation}
C_i = \frac{1}{C_0}\sum_{j=1}^{N} D_j^{n-\alpha} 
\exp{\left(-4 \ln{2}  \left[   \frac{\Delta x_j^{\prime 2} }{b_{\rm maj}^ 2}  +  \frac{\Delta y_j^{\prime 2} }{b_{\rm min}^ 2} \right]\right)},    \label{correction}
\end{equation}
where the factor $C_0$ stands for normalization to unity at the core origin.
In practice, the summation is extended up to ejection ages of 500 d, which is a safe limit in our case to ensure that all relevant terms are included.

The result of numerically evaluating Eq. \ref{correction} using the jet geometry of the ALMA observing date
 is shown in Fig. \ref{Cn}. The factor $1/C_i$ to be applied to the observed brightness profile
is plotted for the east and west jet sides.

\begin{figure}
   \centering
   \includegraphics[width=\hsize]{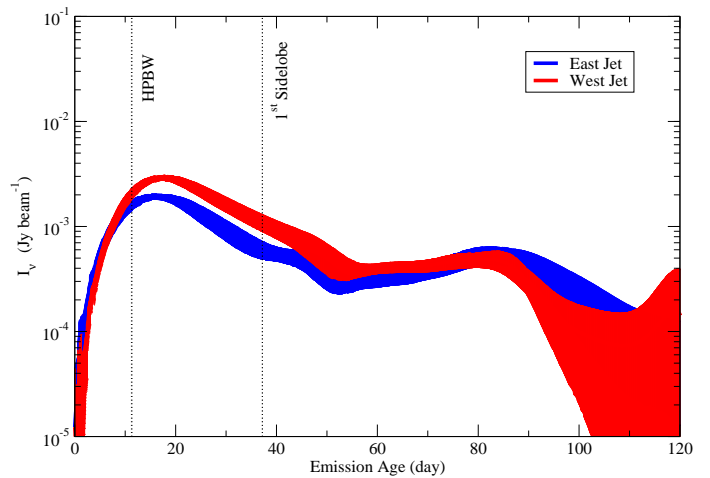}
      \caption{Intrinsic brightness  profile of the SS~433 jets on 2015 September 28 plotted  as in Fig. \ref{profile_obs}, but
      with Doppler boosting and projection effects being corrected.
      The central core has been modelled as a variable point source and fully subtracted in the Fourier plane, which causes the flux density to drop to zero at the origin.}
              \label{profile_varsb}
   \end{figure}

\begin{figure}
   \centering
   \includegraphics[width=\hsize]{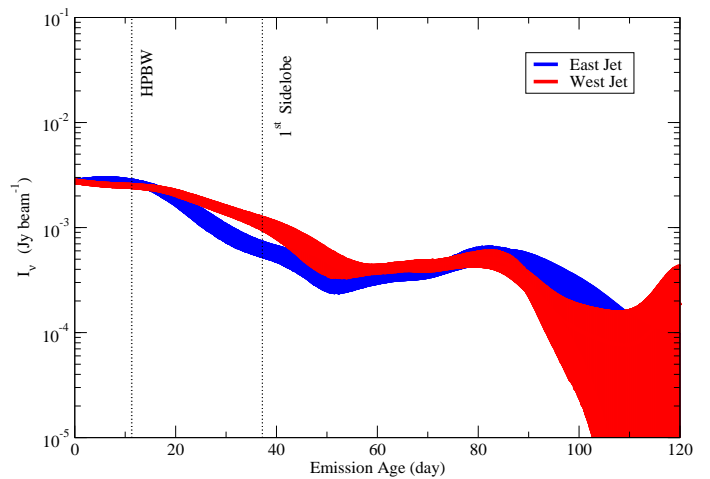}
      \caption{Another estimate of the intrinsic brightness  profile of  the SS~433 jets on 2015 September 28.
      In this figure,  the central core has been modelled  according to the
      brightest CLEAN components in the core vicinity and subtracted in the Fourier plane. This procedure leaves
      some non-zero residual flux density at the origin.}
              \label{profile_uvsub}
   \end{figure}

Before applying the correction, the SS~433 central compact nucleus needs to be removed from the Fig. \ref{profile_obs} data since
our analysis is not meaningful below our angular resolution limits. We have applied two 
 different methods for this purpose. First, 
we removed from the visibility, in the Fourier plane, a variable point source
at the SS~433 core position. Its flux
density was interpolated between the points of  Fig. \ref{lcurve} using an AIPS custom task. The result is shown in Fig. \ref{profile_varsb}.
Second, we removed, also in the Fourier plane, the brightest clean components closer than
the first sidelobe of the synthesized beam from the central SS~433 position. The
AIPS task UVSUB was used 
to perform this procedure. The result is shown in Fig. \ref{profile_uvsub}.
In both cases the result is very similar for 
ages older than $\sim 40$ d. 
The application of the Eq. \ref{correction} correction factors renders the intrinsic brightness profiles of both jets consistent
with being symmetrical, even at millimetre wavelengths.

%

%

\section{Discussion}

The consistence with a symmetric jet brightness profile, evidenced in Figs. \ref{profile_varsb} and \ref{profile_uvsub}
after correcting for projection and beaming effects,
confirms the previous similar findings at centimetre wavelengths   in \citet{2011ApJ...736..118B}. Symmetry, however,
does not mean that the jet brightness profile follows a smooth decay. Although we cannot perform true evolutionary studies,
because only one ALMA observing epoch is available,  we do have some information about the brightness profile as a function of the emission age at the epoch of observation.
Beyond our $\tau \simgt 40$ d threshold, the two profiles in Figs. \ref{profile_varsb} and \ref{profile_uvsub} show a slightly increasing trend for
$55 {\rm ~d}  \simlt \tau \simlt 85 {\rm ~d}$. This behaviour is preceded and followed by intervals of brightness decay for
$45 {\rm ~d}  \simlt \tau \simlt 55 {\rm ~d}$ and $85 {\rm ~d}  \simlt \tau \simlt 120 {\rm ~d}$, respectively. 
Decline, levelling-of,f and decline again is a pattern already observed in the SS~433 jets at centimetre wavelengths, as 
mentioned in \citet{2010ApJ...719.1918R}. The flattening of the decay curve may reflect the effects of previous variability in the core, whose modelling would
require some assumptions on the variable power injection. On the other hand, the decaying sections are more likely related to ageing of the ejected radiating
particles. Assuming that the radio emission is of synchrotron origin, the lifetime $\tau_s$ of a relativistic electron energetic enough to
radiate  at an ALMA frequency $\nu$ under the effects of a magnetic field $B$ is given by
\begin{equation}
\tau_s = 69.4{\rm ~yr} \left[\frac{B}{10{\rm ~mG}}\right]^{-3/2} \left[ \frac{\nu}{233{\rm ~GHz}}\right]^{-1/2}.  \label{vidas}
\end{equation}
This equation is adapted to the mean frequency (233 GHz) of all our ALMA bands.
A first estimate of the magnetic field can be obtained from equipartition arguments using the formalism described
in \citet{1970ranp.book.....P}. The deconvolved angular size of the flat-spectrum radio core in Fig. \ref{map} is about $60 \times 40$ mas$^2$.
The SS~433 average flux density at the central ALMA frequency of 233 GHz was about 88 mJy. This is  within a factor of \simlt 2 from the quiescent emission level
$S_{\nu} = 1.23~{\rm Jy} (\nu/{\rm GHz})^{-0.6\pm0.02}$ estimated by \citet{1982ApJ...260..220S}.
The resulting equipartition field is then 30 mG.

The radiative lifetime of electrons given by Eq. \ref{vidas} is then longer than a decade, while the jet emission in our ALMA map
reaching angular scales up to 1\pri 5 does not exceed an age of about $100$ d. Therefore, synchrotron energy losses do not seem to play
a dominant role, and the faster decay of ALMA emission suggests that another mechanism is at work. 

At this point, expansion losses come as a plausible alternative to explain the overall brightness decay of the SS~433 jets,
as already proposed many years ago in the conical symmetric jet models of radio emitting X-ray binaries developed by \citet{1988ApJ...328..600H}.
Of course, we are excluding eventual brightness enhancement episodes due to the core variability which are beyond the scope of the qualitative assessment intended here.
The semi-analytic \citet{1988ApJ...328..600H} model considers that jet plasma is injected with bulk velocity $v$  at a distance $z_0$ from the core,
 being $r_0$ the jet radius at this point. Relativistic effects are not properly addressed in the context of this model, but
 being SS~433 a mildly relativistic system  this is not a serious drawback for a first-order analysis. In particular, the original model equations use
 a Newtonian time $t$, equivalent to our $t_{\rm age}$ variable, whose difference with $\tau$ is neglected for simple estimative purposes.
 Moreover, the jet is assumed to expand adiabatically in the lateral direction according to $r=r_0 (z/z_0)^p$, where $z$ is the distance from the core
and $p$ a power-law index that parametrizes the expansion rate. Free expansion is achieved when $p=1$, while $p <1$ provides slowed lateral expansion.
We also have $z=z_0 (t/t_0),$ where  $t_0=z_0/v$ is an arbitrary reference time.
Assuming flux conservation, the dominant magnetic field component is the one perpendicular
to the jet axis, scaling as $B = B_0 (r/r_0)^{-1} =B_0 (z/z_0)^{-p}$.
For simplicity, the helical structure of the SS~433 jets is deliberately ignored. 

To estimate whether the jet extent, as a function of frequency, agrees or not with 
adiabatic expansion dominating over radiative losses 
under the above assumptions, we rewrite the optically thin brightness profile 
 given in Eq. (14) of \citet{1988ApJ...328..600H} as
\begin{equation}
\left(\frac{d S_{\nu}}{d z} \right)  = {\rm constants} \times  \nu^{-\frac{(\gamma-1)}{2}}  \left(\frac{z}{z_0}\right)^{-\frac{(7\gamma-1)}{(6 + 6\delta)}}.  \label{profile}
\end{equation}
This approximation is appropriate at large distances from the central source where the jet optical depth is expected to drop well below unity.
Here $S_{\nu}$ is the local jet flux density in this optically thin regime of synchrotron emission.
We also define the radio spectral index $\alpha$ as $S_{\nu} \propto \nu^{\alpha}$. From synchrotron theory \citep{1970ranp.book.....P}
$\alpha=(1-\gamma)/2$, where $\gamma$ is the power-law index of the energy distribution of relativistic electrons.
On the other hand, the constant $\delta$ is equal to 0 or 1, equivalent to the representative cases of free  and slowed expansion  (with $p=1$ or $p=1/2$, respectively).

   \begin{figure}
   \centering
   \includegraphics[width=\hsize]{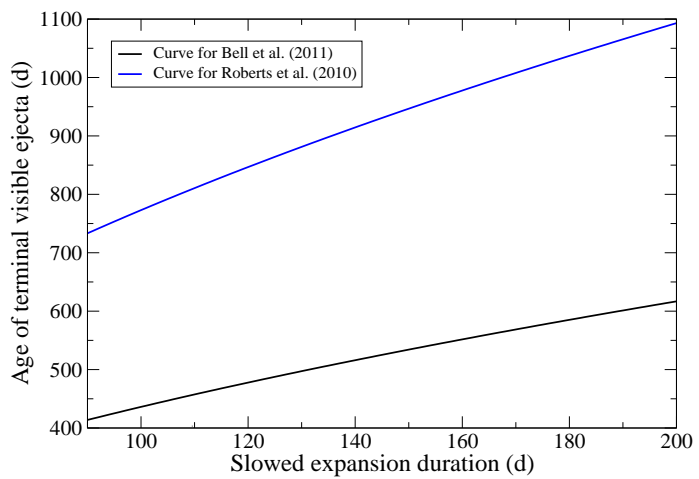}
      \caption{Age of the oldest visible ejecta in VLA C-band (5 GHz) published maps consistent with the observed jet properties in ALMA observations centred at 233 GHz
      according to Eq. \ref{t2free}. Different values of the slowed expansion duration are assumed.}
              \label{tslowfig}
   \end{figure}
%

At our average ALMA frequency ($\nu_1 \simeq 233$ GHz), the oldest ejecta visible ($4\sigma$ level)  has an age $t_1\simeq 90$ d and 
has travelled as far away as $z_1 / D \simeq$1\pri 5. How does this compare with observations at lower frequencies?
Let us assume that we can sample the jet up to some minimum threshold values $(d S_{\nu_1}/d z)_{\rm min}$ and $(d S_{\nu_2}/d z)_{\rm min}$, 
both dependent on instrumental sensitivity.
Considering Eq. \ref{profile} for two frequencies to be compared (here $\nu_2 < \nu_1$),  and taking their equation ratio, we obtain
\begin{equation}
t_2 = t_1 \left[\frac{(d S_{\nu_1}/d z)_{\rm min}}{(d S_{\nu_2}/d z)_{\rm min}}\right]^{\frac{(6+6\delta)}{(7\gamma-1)}}
\left[\frac{\nu_1}{\nu_2}\right]^{\frac{(3+3\delta)(\gamma-1)}{(7\gamma-1)}}.  \label{t2slow}
\end{equation}
The ratio of minimum profile values,  both at the same significance level, 
is in the same proportion as the quotient of the  respective rms noises in the different frequency maps. 
 Here we are comparing our ALMA observations with the 
 \citet{2011ApJ...736..118B} 
 and \citet{2010ApJ...719.1918R}
 VLA maps at centimetre wavelengths (C-band, $\nu_2 = 5$ GHz).  The corresponding rms noise values are  32, 40, and 13 $\mu$Jy beam$^{-1}$.
Scaling by the different beam solid angles,  these numbers are equivalent to 1332, 357, and 75 $\mu$Jy arcsecond$^{-2}$, respectively.
 The optically thin spectral index is taken
 as $\alpha=-0.74$ from the same multi-frequency data of \citet{2011ApJ...736..118B}; hence $\gamma=2.48$.
 
 Now
 we can tentatively explore different options with the values stated above. If the jet expansion proceeds in a slowed way ($\delta=1$), Eq. \ref{t2slow} 
 predicts that the two C-band observations with the VLA should reveal the jets up to $t_2 \simeq1900$ d and $\sim6000$ d, that is,
 more than 5 and 16 years after ejection ($z_2 / D \simeq 15^{\prime\prime}$-$45^{\prime\prime}$).
Clearly this is hard to reconcile with the \citet{2011ApJ...736..118B} 
 and \citet{2010ApJ...719.1918R} images where the oldest ejecta is about 500 d and 800 d, or 1.4 and 2.2 years  old, extending over only a few arcseconds.
 We recall here that \citet{1988ApJ...328..600H} pointed out the possibility that jet expansion could transition from a slowed to a free stage some time
 after ejection.  A difference in the ambient conditions could be a conceivable physical reason behind this change. Intuitively, faster free expansion   enhances the
 adiabatic energy losses of the radiating electrons, thus leading to an earlier brightness decay  compared to slowed expansion. To quantify this effect,
 let $t_{\rm slow}$ be the maximum duration of the slowed expansion phase. 
 Beyond $z_{\rm slow} = v t_{\rm slow}$, or $t\geq t_{\rm slow}$,  the jet profile evolution given by Eq. \ref{profile} 
 must begin to evolve with a $\delta=0$ behaviour. The corresponding expression that ensures continuity is
 \begin{equation}
 \left(\frac{d S_{\nu}}{d z} \right)  = {\rm constants} \times  \nu^{-\frac{(\gamma-1)}{2}}  \left(\frac{z_{\rm slow}}{z_0}\right)^{-\frac{(7\gamma-1)}{(6 + 6\delta)}}
 \left( \frac{z}{z_{\rm slow}} \right)^{-\frac{(7\gamma -1)}{6}}.  \label{profile_free}
 \end{equation}
 Now if the terminal visible ejecta at  frequencies $\nu_1$ and $\nu_2$
 are under the regimes of Eqs. \ref{profile} and \ref{profile_free}, and we take  their ratio again, their respective ages  relate according to
 \begin{equation}
 t_2 = t_{\rm slow} \left(  \frac{(d S_{\nu_1}/d z)_{\rm min}}{(d S_{\nu_2}/d z)_{\rm min}}\right)^{\frac{6}{(7\gamma-1)}}      
            \left(  \frac{\nu_1}{\nu_2}   \right)^{\frac{3(\gamma-1)}{(7\gamma-1)}}
            \left( \frac{t_1}{t_{\rm slow}}   \right)^{\frac{1}{(1+\delta)}}. \label{t2free}
 \end{equation}
 The threshold profile values are also to be replaced by the previous rms noise values per squared arcsecond.
 The  Eq. \ref{t2free} predictions have been plotted in Fig. \ref{tslowfig} for different values of $t_{\rm slow}$ ranging from $t_1=90$ d onwards.
 As we can see, getting the faintest C-band visible ejecta with an age of about 500 d and 800 d  is feasible,
 but it requires an initial  slowed expansion phase lasting for $t_{\rm slow} \simeq 110$ and $130$ d before free expansion gets in.
 The consistency of the two $t_{\rm slow}$ estimates, both pointing to a four-month interval,  is really remarkable given the very different sensitivities of the VLA observations involved.
 The ALMA jets can no longer be traced  during the following unimpeded expansion regime.
  
 Adiabatic losses thus appear  to be a plausible effect that can help us  understand 
 the overall radio properties of SS~433 from centimetre to millimetre wavelengths.
The details of the jet evolution require a modelling work that is more sophisticated than our simple analysis aimed at identifying
the most relevant energetic effects.

\section{Conclusions}

We have reported the results of archival ALMA observations of the microquasar SS~433 that provide a
resolved view of its precessing radio jets at millimetre wavelengths. In addition to this new image,
our main findings can be summarized as follows:

\begin{enumerate}

\item The central core is highly variable  with $\pm 20$\% amplitude flares in the millimetre domain,  evolving on $\sim10^3$ s timescales.
Its spectral index is nearly flat with hints of steepening towards
optically thin values at larger angular distances. This is in agreement with previous observations at lower radio frequencies. A magnetic field value of 30 mG is
found from equipartition arguments given the core  deconvolved angular size of tens of mas.

\item The observed jet path, as imaged with ALMA,
agrees remarkably well with the kinematic model prediction using the  \citet{ 2001ApJ...561.1027E} parameters.
Emission from both jets of SS~433  is detected up to $\sim$1\pri 5 from the central core. This is equivalent
to emission ages of about 90 d for the oldest jet components detected.

\item The jet flow appears to be symmetric after accounting for Doppler boosting and projection effects. 
The brightness profiles of the two jets have an overall decaying trend, although not in a strictly monotonic way.
This is also consistent with previous images in the centimetre domain.

\item Radiative synchrotron losses are not very relevant for the energetic evolution of the radio emitting particles.
In contrast, we confirm the early results in the work of \citet{1988ApJ...328..600H} where adiabatic expansion losses
played a dominant role. Moreover, the jet expansion needs to proceed at a slowed pace for about four months after ejection before a free expansion regime is achieved.
Only in this way are we  able to simultaneously account for the different radiative lifetimes derived from the overall jet properties in ALMA and VLA radio images.
 
\end{enumerate}

\begin{acknowledgements}

This paper makes use of the following ALMA data: ADS/JAO.ALMA\#2013.1.01369.S. ALMA is a partnership of ESO (representing its member states), NSF (USA) and NINS (Japan), together with NRC (Canada) and NSC and ASIAA (Taiwan) and KASI (Republic of Korea), in cooperation with the Republic of Chile. The Joint ALMA Observatory is operated by ESO, AUI/NRAO, and NAOJ.
The National Radio Astronomy Observatory is a facility of the National Science Foundation operated under cooperative agreement by Associated Universities, Inc.
This work was supported by the Agencia Estatal de Investigaci\'on grants AYA2016-76012-C3-1-P and AYA2016-76012-C3-3-P from the Spanish Ministerio de Econom\'{\i}a y Competitividad (MINECO); by the Consejer\'{\i}aa de Econom\'{\i}a, Innovaci\'on, Ciencia y Empleo of Junta de Andaluc\'{\i}a under research group FQM-322; by grant MDM-2014-0369 of the ICCUB (Unidad de Excelencia `Mar\'{\i}a de Maeztu'); and by the Catalan DEC grant 2014 SGR 86, as well as FEDER funds. 

     \end{acknowledgements}

%
%

\bibliographystyle{aa} 
\bibliography{references.bib} 

%

\end{document}